# DL4Proteins Jupyter Notebooks Teach how to use Artificial Intelligence for Biomolecular Structure Prediction and Design

Michael Chungyoun[1], Gabe Au[2] , Britnie Carpentier[1] , Sreevarsha Puvada[1] , Courtney Thomas[3], Jeffrey J. Gray[1,2]

[1] Department of Chemical and Biomolecular Engineering, Johns Hopkins University, Baltimore, MD, 21287, USA.

[2] Program in Molecular Biophysics, Institute for Nanobiotechnology, and Center for Computational Biology, Johns Hopkins University, Baltimore, MD, 21287, USA.

[3] Program in Bioinformatics, Johns Hopkins University, Baltimore, MD, 21287, USA.

## 1 Abstract

Computational methods for predicting and designing biomolecular structures are increasingly powerful. While previous approaches relied on physics-based modeling, modern tools, such as AlphaFold2 in CASP14, leverage artificial intelligence (AI) to achieve significantly improved performance. The growing impact of AI-based tools in protein science necessitates enhanced educational materials that improve AI literacy among both established scientists seeking to deepen their expertise and new researchers entering the field. To address this need, we developed DL4Proteins, a series of ten interactive notebook modules that introduce fundamental machine learning (ML) concepts, guide users through training ML models for protein-related tasks, and ultimately present cutting-edge protein structure prediction and design pipelines. With nothing more than a web browser, learners can now access state-of-the-art computational tools employed by professional protein engineers - ranging from all-atom protein design to fine-tuning protein language models for biophysically relevant functional tasks. By increasing accessibility, this notebook series broadens participation in AI-driven protein research. The complete notebook series is publicly available at https://github.com/Graylab/DL4Proteins-notebooks.

## 2 Introduction

The 2024 Nobel Prize in Chemistry was awarded to David Baker for his groundbreaking contributions in computational protein design, and to Demis Hassabis and John Jumper for their breakthrough in protein structure prediction. The award celebrates achievements powered by innovative deep learning (DL) methods. These advancements have enabled the creation of medicines that outperform natural counterparts in addressing both well-studied[1–7] and understudied[8] disease, and proteins with novel functions beyond those found in nature[9], offering promising solutions in fields from materials and nanotechnology[9–15] to environmental issues[16–18]. The profound impact of AI extends beyond protein design, as evidenced by the Nobel Prize in Physics of the same year awarded to John J. Hopfield[19] and Geoffrey Hinton[20] for their foundational discoveries in ML with artificial neural networks. In today's landscape, AI literacy is essential for success in the field of protein design and prediction.

## 2.1 Scientific background

Prior to the rise of DL, computational modeling of biomolecular structure and function often used (1) dynamic simulation tools or (2) sequence- and structure-based energy minimization tools. Molecular dynamics (MD) simulations model the time-dependent behavior of molecular systems by numerically integrating Newton's equations of motion, enabling the prediction of structural, thermodynamic, and kinetic properties based on interatomic forces and energy functions. MD is widely used in computational biology and chemistry and is implemented in software packages such as GROMACS[21], AMBER[22], NAMD[23], LAMMPS[24], Desmond[25], MOE[26], and CHARMM[27].

Energy minimization tools identify dominant conformational states relevant under biological conditions. These methods often accelerate computation using approximations such as fixed bond lengths and angles, implicit solvent models, and empirically derived energy functions. Tools like Rosetta3[28,29] and FoldX[30] can model large biomolecules and complexes at equilibrium without long-timescale simulations. Rosetta uses a composite energy function (a linear combination of physical, empirical, and statistical terms) to model biophysical characteristics like packing, solvation, and torsional preferences[31]. It supports structure prediction, docking, and design for proteins, nucleic acids, and carbohydrates[32]. A Python interface, PyRosetta, provides programmatic access to its algorithms[33].

Rosetta has long been a top performer in blind protein structure prediction challenges. Then, in CASP14[34], AlphaFold2[35]— a DL method developed by DeepMind—set a record by significantly outperforming Rosetta and other leading approaches. On average, its predictions were twice as accurate as the next best models and it consistently achieved near-experimental accuracy across all target types, with a Global Distance Test Total Score (*GDT TS**) exceeding 90%[34]. AlphaFold2's success stems from its use of advanced DL techniques, including a transformer-based architecture with attention mechanisms and equivariant neural networks, enabling it to capture the complex rules of protein folding with unprecedented precision.

DL models rely on deep neural networks, which are layered structures that learn complex patterns by adjusting weighted connections through data-driven optimization. Biomolecules can be represented to neural networks in diverse ways such as graphs, sequences, or 3D voxel grids, enabling models to learn underlying data distributions rather than relying on hand-crafted physical approximations. This shift allows deep networks to capture complex, nonlinear patterns across a range of tasks, from structure prediction and molecular docking to protein design and dynamics[36]. A wide array of architectures, including transformers[37], graph neural networks[38], and diffusion models[39] discover solutions to specific biological problems or capture general principles to repurpose across many tasks. The emergence of large, pre-trained models, continual refinement of DL architectures, and increased availability of protein sequence, structure, and function data is accelerating progress, offering increasingly generalizable and high-performing solutions across the biomolecular space.

## 2.2 Pedagogical background

There are several challenges inherent with teaching AI in protein design. First, the field is extremely multidisciplinary. Understanding ML requires knowledge in linear algebra, calculus, and statistics; using ML models and their operations requires knowledge in object-oriented programming; and understanding the use cases in biology requires knowledge in the building blocks of biomolecular structures and the biophysical interactions involved. Students typically have varying levels of knowledge in each of the disciplines. Second, training and utilizing large DL models require powerful hardware such as graphics processing units (GPUs) and specialized processors optimized for parallel computation. These GPUs are costly and may be inaccessible to many students. Finally, mastering this space requires hands-on experimentation with a wide array of models. This often involves managing software dependencies, which can be a significant barrier for students with limited programming experience. Without accessible tools or support, these challenges may discourage otherwise capable learners from engaging with the field.

Educational material for teaching physics-based computational protein structure prediction and design exists. Molecular dynamics platforms often include accompanying beginner tutorials for their usage. CHARMM-GUI includes lessons for running simple simulations[40], while the GROMACS platform has tutorials from simple lysozyme simulations to more complicated simulations, such as umbrella sampling[41]. MOE has been used in engineering curricula and has an online cookbook for various applications from molecular docking to homology modeling[42]. FoldIT[43] and EteRNA[44] held week-long programs and published curricula for undergraduate and high school students to learn principles of protein and RNA biochemistry[45]. The RosettaCommons developed a PyRosetta notebook series and an accompanying YouTube video lecture series[46] to teach the scientific capabilities of Rosetta (sampling and scoring biomolecular conformations) and present advanced applications (antibody design, membrane protein modeling, and RNA structure prediction)[47]. The Meiler lab offers workshops on protein-protein and ligand docking, design with non-canonical amino acids, enzyme design, and antibody design using the BioChemical Library and Rosetta[48]. Schrodinger provides paid online certification tutorials for industry and academic scientists to learn about computational antibody engineering, molecular/quantum mechanics, free energy calculations, ligand docking, and other uses of their tools for therapeutic development[49]. The ML4Bio workshop focuses on teaching how to identify problems in computational biology, the main parts of a typical ML workflow, how to compare specific classifiers, perform model selection, evaluate a model on new data, and judge the use of ML in biological contexts[50].

There are relatively few comprehensive resources focused on applying AI across the diverse spectrum of protein design and prediction tasks. Aspiring scientists and established researchers alike often face significant barriers when seeking to learn about AI tools for protein design, as they must navigate the supplemental sections of complex, jargon-laden research papers. Many research papers are not written to teach–they're instead written in a way to assert novel claims, meanwhile assuming readers have all the context. Such documents can be inaccessible and intimidating, particularly for those with expertise in biology but limited exposure to AI. While some publications provide example scripts, engaging with these resources requires overcoming several challenges: an understanding of machine learning concepts to interpret model functions, programming proficiency to troubleshoot code and manage environmental dependencies, and access to high-performance computational resources, such as GPUs, for running inference.

## 2.3 Target audience and purpose

Our notebook series is tailored to the needs of three broad categories of learners:

1. Upcoming scientists in their undergraduate or graduate education who seek to learn about AI applied to proteins towards entering this career space after school;
2. Professionals with experience primarily in the wet lab side of protein design who recognize the significance of AI and seek to learn how to apply these tools in their workflows; and
3. Educators who seek to learn how to effectively teach this material in their courses.

These diverse learners bring varying levels of academic and professional experience, computational skills, and institutional support for computational education and resources. However, they share a common interest in biomolecular design and AI; they value clear and structured explanations supplemented by visual aids, workflows, and hands-on examples; and they are committed to investing time to learn about AI tools for proteins. Moreover, they desire low-barrier, user-friendly entry points for those with minimal coding or machine learning experience.

The objectives of our workshop series are to:

1. Provide learners with a foundation in neural networks and ML, fostering sufficient intuition about the mathematics behind AI models.
2. Equip scientists with advanced tools for biomolecular structure prediction and design, enhancing their proficiency in utilizing state-of-the-art technologies; and
3. Prepare new students to enter the AI and biomolecular design fields as competent scientists by integrating AI and biology education.

Our overarching goal is to incorporate cutting-edge AI tools into protein design education. We advocate for structured instruction in biomolecular design and promote the broader adoption of AI in protein design curricula through our notebook series as a model for future educational frameworks. This approach will prepare the next generation of leaders in AI and biomolecular structure prediction and design, accelerating both the development of current molecular solutions and the discovery of entirely new applications.

# 3 Results

## 3.1 Overview of notebook topics

Each notebook within DL4Proteins follows a similar structure: an objectives section outlining key concepts learners should be able to define; the main technical content interspersed with active learning questions; a set of longer-form exercises that may require revisiting and modifying earlier code; and a concluding list of relevant AI for protein design papers for further reading. The notebook series progresses from (1) fundamental ML concepts (neural network basics, PyTorch fundamentals, convolutional neural networks) to (2) training and applying models to toy datasets and proteins (language models for generating first Shakespeare-like text and then protein sequences, graph neural networks on protein structures, diffusion models on sine waves) to (3) advanced pipelines of modern protein design tools (fine-tuning protein language models, AlphaFold for structure prediction, end-to-end *de novo* protein design pipeline, all-atom protein design with a diffusion model) (**Fig. 1**). After completing the workshop series, learners will be able to explain the fundamental concepts and operations that occur under the hood of a ML model, train ML models for a variety of tasks, and use real world examples of computational tools that mimic pipelines used in current bleeding edge research papers. The overall objectives are provided in **Table 1,** and the detailed objectives of each workshop are provided in **Table2**.

### 3.1.1 Introduction to machine learning concepts

Part I introduces basic neural network concepts using both NumPy and PyTorch libraries to lay a foundation in ML techniques before moving to domain-specific applications (**Fig. 2**).

**Notebook 1: Neural Networks with NumPy** provides a detailed, hands-on experience with neural networks by building a neural network from scratch using matrices and vectors (in NumPy), rather than relying on high-level DL libraries. The topics include how neurons work, how information moves forward and backward through a network, how activation functions (like ReLU and softmax) shape outputs, how loss is calculated, how the network learns through backpropagation, how optimizers adjust the model, and the overall steps of training a neural network, culminating in practical coding examples and visualizations to reinforce learning.

**Notebook 2: Neural Networks with PyTorch** provides learners with hands-on experience using the PyTorch classes while reinforcing the neural network concepts introduced in the previous notebook. It covers neural network instantiation and definition, execution of a training loop for a classification task, and performance visualization, all while emphasizing the efficiency of PyTorch operations. The workshop includes active learning questions to help learners understand how modifications to the neural network's components affect its performance (example in **Fig. 3**).

**Notebook 3: Convolutional Neural Networks** focuses on the distinction between convolutional neural networks (CNNs) and dense neural networks by exploring the operations performed in convolutional layers and how these layers learn and represent features from input images. The notebook introduces the convolution concept through simple matrix operations, progresses to 2D convolution with real MNIST images, and includes active learning questions to enhance understanding of CNN architecture and training processes, all while providing visual comparisons and code implementations for clarity.

Together, these foundational workshops provide participants an overview of the ingredients in a basic deep neural network model and prepare learners to understand how elements combine in more complex neural network architectures ahead.

### 3.1.2 Training deep learning architectures

Part II focuses on training transformer-based language models, graph neural networks (GNNs), and denoising diffusion probabilistic models (DDPMs), adapted specifically for protein sequence and structure data.

Language models (LMs) trained on natural language sequences are designed to reason and respond to prompts from user input[51–55]. Protein LMs are trained on protein sequences represented by strings of amino acids and utilized to design novel protein sequences or predict protein function[56]. Prominent examples include ESM3[57] and Progen3[58], which have been used to generate novel, functional proteins. In **Notebook 4: Language Models for Shakespeare and Proteins**, students train an autoregressive LM from scratch, beginning with a natural language dataset (Tiny Shakespeare[59]) to illustrate the fundamental operations of next-token prediction and tokenization before progressing to protein sequences, which can be more complex. The notebook covers dataset preparation, the autoregressive training scheme, instantiation and training of the LM on both English and protein languages, while also discussing the significance of language models like ChatGPT[60] and the role of attention mechanisms. Active learning questions allow learners to observe and report how changing various parameters of the model's size impacts how well the model LM from the textual dataset.

Graph Neural Networks (GNNs)[38] offer a powerful framework for modeling protein structures by representing amino acid residues as nodes (each carrying features such as residue identity, sequence position, or local environment) and as edges (a representation of the relationship between residues) to capture spatial or biochemical interactions between residues. GNNs respect equivariance or invariance to rotations and translations allowing them to model protein geometry in a way that reflects the physical reality of molecules in 3D space. Recent methods such as ProteinMPNN[61] and ESM-IF[62] demonstrate the success of GNN-based models in tasks like sequence design and structure prediction, showcasing the potential of these networks to capture both local and global structural features in proteins. **Notebook 7: Graph Neural Networks** teaches graph theory and its application to neural networks by representing molecular structures as graphs, with atoms as nodes and bonds as edges, or connections between nodes. The notebook covers essential concepts such as types of data suitable for GNNs, the differences between directed and undirected edges, the principles of message passing, and how GNN layers operate[63]. The notebook provides active learning questions, visual representations of graph structures, and guidance on preparing protein data for GNN input, culminating in the implementation and evaluation of a Graph Convolutional Network (GCN) model[64,65].

Owing to their heuristic flexibility, diffusion models offer a powerful framework for the generation and design of diverse biomolecular structures, including proteins, nucleic acids, and small molecules. Their capacity to learn complex data distributions enables both broad exploration across molecular space and targeted sampling from sub distributions characterized by specific structural or functional properties [66]. **Notebook 8: Denoising Diffusion Probabilistic Models (DDPMs)** explores the principles and implementation of diffusion models, which are generative models that create data samples from noise, providing superior generation quality and lower computational costs compared to traditional models like generative adversarial networks and variational autoencoders. The notebook details the two key components of diffusion models: the forward process that gradually adds noise to data during training and the reverse process that learns to reconstruct the original data, while covering the mathematical foundations, loss interpretation, and practical implementation of these processes in code.

### 3.1.3 Pipelines for protein structure prediction and design tasks

The foundational ML concepts provided in part I combined with the development of basic DL architectures in part II prepare learners to engage with state-of-the-art biomolecular structure prediction and design tools in part III. These notebooks highlight the methodological innovations behind these models and demonstrate how to apply them in practice.

**Notebook 5: *Language Models for Transfer Learning*** introduces fine-tuning techniques, i.e., methods that adapt pre-trained models to new, domain-specific tasks by updating their parameters on specialized datasets. Fine-tuned protein language models play a key role in pre-clinical drug discovery campaigns, demonstrating effectiveness in predicting therapeutic and developability properties of antibodies such as thermostability[67,68], expression[69], aggregation[69], pharmacokinetics[70], immunogenicity[71], polyreactivity[69,72,73], binding affinity[74–76], and viscosity[72]. In this module, fine-tuning is applied to classify residues in protein binding sites and to use embeddings from the language model for ligand binding site prediction. The notebook additionally addresses the challenges of fine-tuning (such as overfitting in low data regimes) and then explores a more effective approach using Low-Rank Adaptation (LoRA), visually comparing performance outcomes and providing resources for further study[56,77].

AI-based protein structure prediction models have vastly increased the accessibility of accurate protein models, offering a rapid alternative to experimental structural biology, which can take months to solve a single structure and cannot match the pace of the growth of the already known protein sequences. While the Protein Data Bank contains 200,000 resolved structures, the AlphaFold Database[78] covers over 200 million UniProt sequences and the ESM Metagenomic Atlas[79] provides 600 million metagenomic sequences. **Notebook 6: Introduction to AlphaFold** takes a practical, no-coding approach to explore how the AlphaFold model functions as a “black box”. The notebook engages learners with interactive predicted protein folds (**Fig. 3**), provides descriptions of model inputs and outputs, and guides how to interpret key performance metrics provided by the model like coverage and pLDDT, and concludes with resources for further exploration of related models in protein structure prediction.

**Notebook 9: Designing Proteins** introduces computational methods for *de novo* protein design, including unconditional protein structure generation, motif scaffolding, and creating novel binders. The notebook outlines a pipeline utilizing RFDiffusion to create novel protein backbones, ProteinMPNN for predicting amino acid sequences that fit given structures, and AlphaFold2 to verify the folding of the designed sequences. The notebook guides learners through the processes of generating proteins, extending existing motifs, and designing new binders for potential therapeutic applications. This pipeline mirrors those used by experts in the field, as RFDiffusion has been used for many successful applications including the neutralization of venom toxins[2] and design of binders for the challenging TNF receptor 1[1].

**Notebook 10: RFDiffusion All-Atom (RFDiff-AA)** extends the design of novel proteins to those capable of binding molecules beyond proteins, leveraging a multi-track architecture that integrates amino acids, nucleic acids, and small molecules for all-atom design. The notebook guides learners through the process of designing new protein sequences for small molecule binding, supplemented by active learning questions to reinforce understanding of the model's architecture and its encoding methods. RFDiff-AA was previously used to *de novo* design diverse, high-affinity, thermally stable proteins for small molecules like digoxigenin, heme, and bilin, achieving functional binding[80].

## 3.2 Strategies to facilitate learning outcomes in notebooks

Teaching AI-based biomolecular structure prediction and design presents cognitive, pedagogical, and resource challenges which we specifically address with the content, teaching style, and programming environment of the notebook series.

AI for protein design presents a cognitive challenge, as many learners may come in with limited or no machine learning experience. The notebook series begins with the building block elements of neural networks, and then progressively increases the complexity of the material until learners reach powerful protein AI tools. To facilitate learning, we present the material in a multimodal format with images, animations, code, and text to enhance student's mental representations of concepts (**Fig.s 4, 5**). Active learning questions in the form of fill-in-the-blank coding sections and open-ended questions give students opportunities to build intuition about how a model's inputs and design choices shape its performance on protein tasks. Using Python allows combining with other scientific code libraries (NumPy, Pandas, Scikit-Learn) and biological packages (PyRosetta, BioPython). PyTorch is extensively documented and there are existing forums for troubleshooting PyTorch-based tasks, increasing the accessibility of ML to non-experts[81]. We also cite materials for deeper learning of PyTorch including YouTube video tutorials, research articles, and blogs[82]. Concerning time and attention constraints for learning, each notebook can be completed in 1.5 to 2 hours, which fits well within the typical college class length, as well as the amount of time we can hold our attention on learning a new task[83]. In some application notebooks, we present only the main topics with graphics while hiding the code in markdown cells. This helps learners who want to apply the models quickly to their own use cases, without delving into the underlying machine learning details. Instead, they can focus on interpreting and understanding the model's output metrics. For instance, in Notebook 6 on AlphaFold, students simply input a sequence to predict a protein structure and then build intuition by interpreting MSA coverage, pLDDT, and PAE values in relation to the predicted structure.

The material in AI for protein design is inherently multidisciplinary. We discretize ML concepts, applications, and pipelines into different notebooks, as shown in **Fig 2**. This allows learners to clearly identify what areas they might focus on given their expertise in ML. For those that are interested in working through all notebooks, we make sure to cover only the most important core ML/DL material so that they can efficiently gain mathematical intuition on the foundations of ML models and then move on to applications for biomolecular structure prediction and design.

Running and training large DL models requires access to computational processing resources, creating a resource challenge. We utilize Google Colab as it offers significant advantages for training and inference, including access to free GPUs, such as an NVIDIA Tesla T4 with 15 GB of memory, and substantial system specs like 13 GB of RAM and 110 GB of disk space. With a 2.25 GHz CPU and continuous sessions lasting up to 12 hours, it provides a robust and flexible environment for running machine learning workflows. Google Colab's ability to handle packages discreetly within a single notebook eliminates dependency conflicts, making it easy to work with a notebook-per-model approach. This enables learners to bypass the need for High-Performance Computing (HPC) resources, as all notebooks can be completed in 1.5 hours and stay within the GPU resource allocation, making it ideal for education.

## 3.3 Learning outcomes: Results from teaching our material

We piloted use of these notebooks in a course entitled "Computational Protein Structure Prediction and Design" at Johns Hopkins University in Spring 2025. Students matriculated with varying amounts of programming experience; some having only taken an introductory programming class while others had already completed graduate level machine learning classes. Regardless, by the end of the semester, students had learned a baseline knowledge of ML and how to use several prominent protein structure prediction and design models. Students spent the second half of the semester developing a project that incorporates the models used throughout the semester or training their own models from architectures introduced in the class. Groups of 2-3 students completed projects ranging from structurally analyzing the behavior of protein self-assembling colloidal systems with AlphaFold3, to designing proteins to novel DNA or protein targets, and improving the predicted binding affinity of existing therapeutic antibody candidates. The course gave students with minimal programming experience an opportunity to learn and apply real machine learning models to impactful protein design tasks. As one former student stated: "I actually had zero Python proficiency at the beginning of this semester… and once the machine learning topics came in, I was worried about how I would manage. The workshops and notebooks made it so much easier to adjust… I truly found your class so interesting, and it blended organic chemistry, thermodynamics and machine learning concepts in a way that was just as thought-provoking as it was practical. This was easily my favorite course I've ever taken! I will definitely stay close to this field because of it."

# 4 Discussion

AI models have proven to be impactful in the biomolecular structure prediction and design space. The multidisciplinary nature at the intersection of machine learning and biophysics leads to several challenges for students, including access to computational resources for running the models. Here we have described a set of notebooks for learning biomolecular structure prediction and design that can be used in a classroom context or for individual self-study. We highlighted the impact of this material for a graduate/undergraduate course in computational protein design, but the reach of the notebooks has also extended beyond the classroom to summer code school programs like the NSF REU[84] and RAMP[85]. Educators at all levels—high school through graduate—can use and adapt these notebooks to build AI literacy and better prepare students for success in protein design and prediction.

The first half of our notebooks focus on basics of machine learning, PyTorch, and neural networks, which for the last few years have remained relatively stable across improvements in deep learning. The latter half of our notebook series highlights today's state of the art models, which indicates some of the models in our notebooks will be superseded by the models of the future. To address this, DL4Proteins will be a dynamic repository, where additional notebooks will continue to be added as new open-source models are released. Some of the upcoming notebooks include flow[86] and discrete diffusion[87].

By enhancing the accessibility of these tools, we hope to accelerate a transformative moment akin to the advancements driven by open-sourced large language models[88], empowering young scientists to explore innovative applications and pioneer improvement in biomolecular protein design. DL4Proteins equips learners with AI-based methods for biomolecular design, expanding the toolkit of future protein engineers. Solving real-world problems in protein design requires the integration of ideas from multiple domains, and this notebook series will increase the quantity, quality, and diversity of learners who can contribute. By teaching how to apply and develop AI tools, these notebooks drive advances in drug discovery, materials science, nanotechnology, and environmental solutions.

# 5 Figures and Tables

## 5.1 Glossary

Box 1. Glossary of Deep Learning Terms

| | |
|---|---|
| **Activation Function** | A mathematical function that determines the non-linear transformation in a neuron so networks can learn complex patterns[89]. |
| **All-atom Design** | Computational protein design methods that operate at atomic resolution, where sidechain and backbone conformations are modeled with explicit atoms to capture steric and energetic details[90]. |
| **Attention Mechanism** | A method that weighs the importance of different tokens in a sequence when processing each token, enabling a model to efficiently capture context and long-range dependencies. |
| **Autoregressive** | A modeling approach where future values are predicted from past values, such as language models that generate the next token based on previous ones. |
| **Backpropagation** | An algorithm that optimizes neural networks by propagating error signals backward and updating weights to reduce prediction loss[91]. |
| **Convolutional Neural Networks (CNNs)** | A neural network architecture that uses local data filters to detect patterns in data. Widely used in image and sequence analysis[92]. |
| ***De novo* Protein Design** | Engineering of novel proteins not observed in nature that adopt stable and/or functional conformations[93]. |
| **Dense Neural Networks** | A network where each neuron in one layer connects to every neuron in the next, enabling rich feature transformations but with high computational cost. |
| **Diffusion Models/Denoising Diffusion Probabilistic Models (DDPMs)** | A class of generative models that gradually add noise to data and train a model to reverse the process, turning random noise into realistic samples. |
| **Embeddings** | Compact numerical representations of data (e.g., words or proteins) that capture semantic or structural properties. |
| **Equivariance** | A property in which applying a transformation to the input leads to the same transformation in the output (e.g., rotating an image rotates the feature map). |
| **Fine-tuning** | A method of refining a pre-trained, general model with data from a specific sub-domain. |
| **Global Distance Test Total Score (GDT_TS)** | A protein structure prediction accuracy metric that reports the average fraction of $C_\alpha$ atoms in a predicted protein model that can be superimposed within 1, 2, 4, and 8 Å of the corresponding positions in the experimental structure after optimal alignment[94]. |
| **Generative Adversarial Networks (GANs)** | A framework where two neural networks—the generator and the discriminator—are trained in competition, with the generator creating data and the discriminator distinguishing real from generated samples. |
| **Graph Neural Networks (GNNs)** | Architectures that learn representations over graph-structured data by propagating and aggregating information across graph nodes and edges to enable prediction or classification tasks. |
| **Invariance** | A property where outputs remain unchanged even if certain transformations are applied to the input. |
| **Language Models/Transformer-based language models (LMs)** | Models that predict or generate sequences of tokens, such as words or amino acids, with transformers providing attention to capture long-range patterns. |
| **Low-Rank Adaptor (LoRA)** | A fine-tuning method that updates only small low-rank matrices within a model, reducing memory and compute while preserving performance. |
| **Modified National Institute of Standards and Technology (MNIST) database** | A dataset of 70,000 handwritten digit images (0–9), widely used as a benchmark for training and testing image classification models. |
| **Motif Scaffolding** | A protein design method where functional motifs are embedded onto stable structural frameworks to preserve activity. |
| **Multiple Sequence Alignment (MSA)** | An alignment of biological sequences that highlights conserved regions reflecting shared structure, function, or ancestry[95]. |
| **Neural network** | A computational model inspired by the brain, made up of layers of interconnected nodes ("neurons") that learn to recognize patterns and make predictions from data. |
| **NumPy** | A Python library for fast numerical computing with multi-dimensional arrays. |
| **Predicted Local Distance Difference Test (pLDDT)** | A per-residue confidence score (ranging between 0-100) from AlphaFold[35] that estimates the accuracy of local backbone and side-chain geometry. Higher values indicate greater reliability of the predicted structure. |
| **Predicted Aligned Error (PAE)** | PAE (predicted aligned error) measures AlphaFold2's confidence in residue–residue positioning. It is defined as the expected error in the location of residue X, in angstroms (Å), after aligning the predicted structure to residue Y in the true structure. |
| **PyTorch** | A Python library that allows for efficient tensor computation with GPU acceleration to build and train deep neural networks[96]. |
| **Rectified Linear Unit (ReLU)** | A common activation function defined as $f(x) = \max(0, x)$, outputting zero for negative inputs and the input itself for positive inputs. |
| **Softmax** | A function that converts raw scores (logits, akin to relative energies) into a probability distribution by exponentiating and normalizing values. |
| **Tokenization** | The process of splitting input data (such as text or protein sequences) into discrete tokens that serve as the model's basic input units. |
| **Variational Autoencoders (VAEs)** | A generative model that encodes inputs into a probabilistic latent space and decodes them to reconstruct or generate new samples. |

## 5.2 Figures

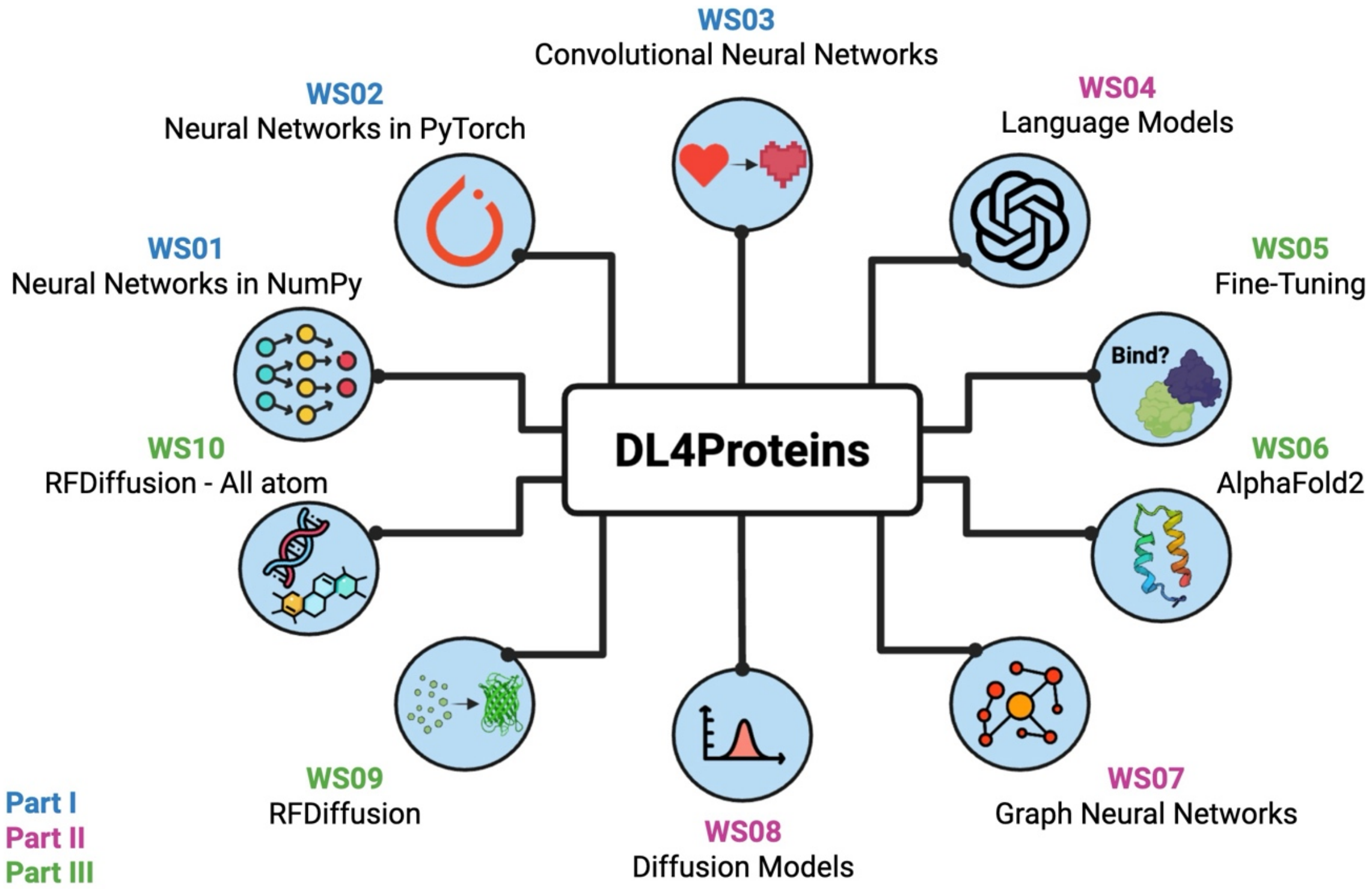

**Figure 1: Notebook topics progress from introductory machine learning to state-of-the-art AI for protein design.** The topics begin with fundamental concepts of machine learning and PyTorch, transitions to training small neural networks for images, languages, and proteins, and then introduces state of the art protein design and structure prediction models. Each workshop title is colored according to whether it falls within Part 1 (ML fundamentals), Part II (DL architectures), or Part III (protein structure prediction and design pipelines).

**Table 1: Summary of overall learning objectives.** The color scheme matches that in Fig. 4.

| Part | Workshops | Learning Objectives |
|---|---|---|
| **Part I: Introduction to Machine Learning** | WS01, WS02, WS03 | • Build and train neural networks using NumPy and PyTorch<br>• Explain and apply general principles behind model architecture, loss, activation functions, and optimization |
| **Part II: Advanced Deep Learning Models** | WS04, WS07, WS08 | • Train and optimize language models and graph neural networks<br>• Implement diffusion models for generative design |
| **Part III: Applied Protein Design Tools** | WS05, WS06, WS09, WS10 | • Finetune pre-trained protein LMs for protein function prediction<br>• Predict protein structures and analyze model predictive performance<br>• Design novel proteins and binders using diffusion models and language models |

**Table 2: Summary of learning objectives for each notebook.** The color scheme matches that in Fig. 4.

| WS Number | WS Name | Topics Covered | Learning Objective |
|---|---|---|---|
| **WS01** | **Neural Networks with NumPy** | Neuron<br>Forward/backward pass<br>Activation function<br>Loss function<br>Optimizer<br>Weights and biases<br>Training and epochs | Interpret and implement neural networks using NumPy; Perform forward and backward propagation calculations; Train simple neural networks and visualize decision boundaries |
| **WS02** | **Neural Networks with PyTorch** | PyTorch literacy<br>Multi-class classification | Build neural networks in PyTorch; Define training loops and loss functions; Visualize learned classification boundaries; Compare architectures and training dynamics |
| **WS03** | **Convolutional Neural Networks** | Convolutions<br>Kernel<br>Convolutional models<br>MNIST | Perform convolution operations; Build CNNs in PyTorch; Train CNNs on image data; Visualize feature maps and filters learned during training |
| **WS04** | **Language Models for Shakespeare and Proteins** | Tokenization<br>Next-token prediction<br>Transformer<br>Attention mechanism<br>GPT | Prepare and analyze natural and protein language datasets; Instantiate transformer-based language models; Train autoregressive models for text and protein sequence generation |
| **WS05** | **Language Model Embeddings and Transfer Learning** | Fine-tuning<br>ESM language model<br>LoRA<br>Binding site prediction | Quantify benefits of finetuning protein LMs; Apply LoRA for efficient fine-tuning; Use embeddings for ligand-binding site prediction |
| **WS06** | **Introduction to AlphaFold2** | AlphaFold2<br>Multiple sequence alignment<br>Template structures | Define role of key portions of architecture: input data (sequence, MSA, templates), Evoformer, and |
| | | MSA coverage<br>pLDDT<br>Predicted aligned error | Structure Module; Predict protein structures using ColabFold; Interpret pLDDT and PAE outputs; Analyze MSA coverage and confidence metrics |
| **WS07** | **Graph Neural Networks for Proteins** | Node attributes<br>Edge attributes<br>Adjacency matrix<br>(Un)directed graphs<br>Message passing<br>GCN | Represent proteins as graphs (nodes/edges); Apply message passing in GNNs; Perform node/graph/edge-level protein predictions; Define invariance/equivariance in GNNs |
| **WS08** | **Denoising Diffusion Probabilistic Models** | Forward diffusion<br>Reverse diffusion<br>Scheduler<br>Diffusion loss<br>Time embeddings<br>Posterior variance | Define main processes involved in diffusion model; Implement forward and reverse diffusion; Explain and compute loss functions like ELBO and KL divergence |
| **WS09** | **End-to-End Protein Design (Diffusion + AlphaFold + ProteinMPNN)** | RFDiffusion<br>ProteinMPNN<br>AlphaFold2<br>Unconditional diffusion<br>Motif scaffolding<br>Binder diffusion | Perform *de novo* protein backbone generation with RFDiffusion; Design sequences with ProteinMPNN; Validate structures with AlphaFold; Integrate design pipeline for practical applications |
| **WS10** | **RFDiffusion All-Atom** | All atom diffusion<br>Multi-track architecture<br>Binder design | Design proteins binding to small molecules using RFDiffusion All-Atom; Apply atomic-level graph representations; Use multi-track models for 3D spatial design; Handle small molecules and covalent modifications in protein design |

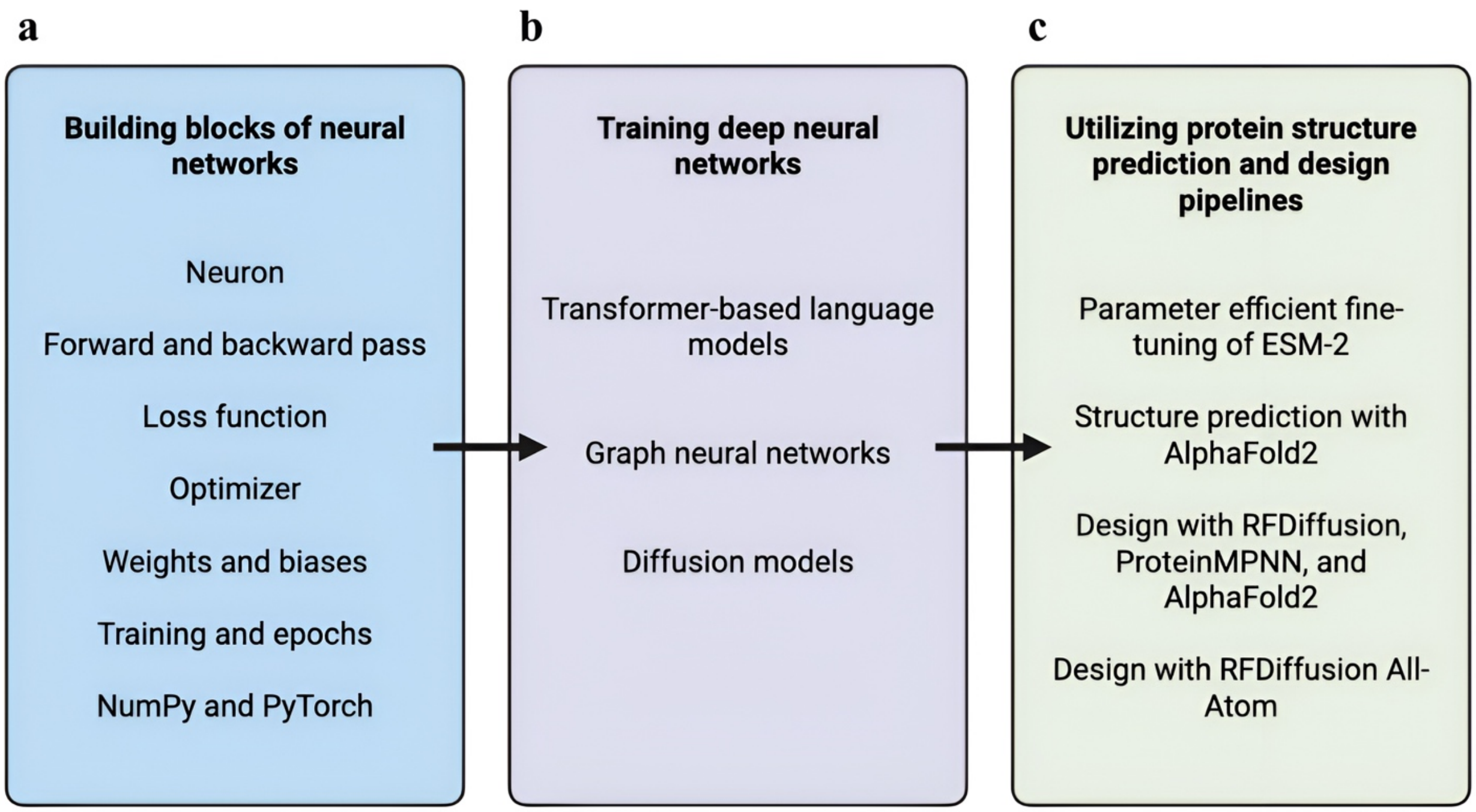

**Figure 2: Notebooks are categorized based on whether they cover ML basics, training of deep learning architectures, or focus on a protein engineering pipeline**. We progressively increase the difficulty of the material, starting with basic concepts of neural networks in a) part I (blue), to how those concepts come together when training a deep neural network in b) part II (purple), and finally pre-trained state of the art AI models for protein structure prediction and design in c) part III (yellow).

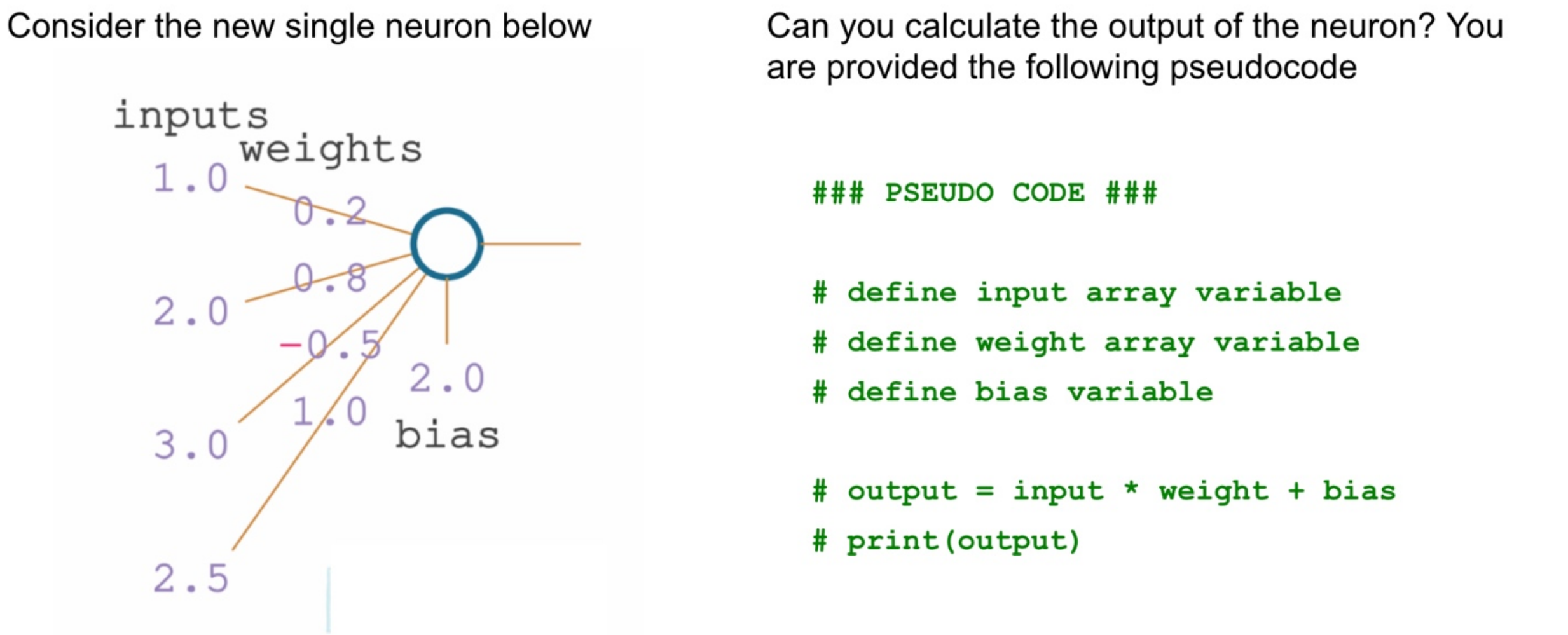

**Figure 3: Active learning questions are provided in each notebook to engage learners and provide an opportunity for immediate feedback.** In the following example active learning question from “Workshop 1: Neural Networks with NumPy”, learners must compute the output of the neuron themselves by applying the concept of the neuron equation. The problem is presented in multiple formats (neuron diagram and pseudocode) and the structure invites students to build an intuition on how neurons behave.

### Instantiating model

We will be using a feedforward neural network consisting of two dense (fully connected) layers with ReLU activation, where the first layer takes 2 input features and outputs 64 neurons, and the second layer takes 64 neurons and outputs 3 neurons.

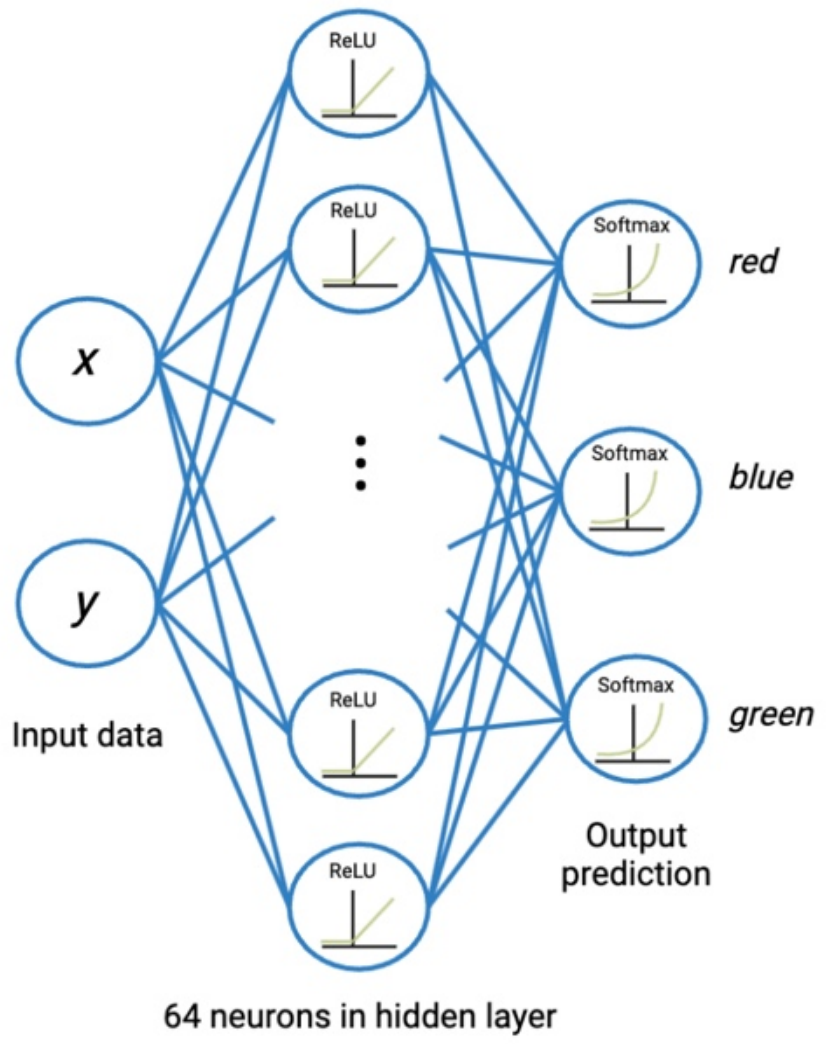

**Loss**
Categorical cross entropy

$$L(y, \hat{y}) = -\frac{1}{N} \sum_{i=1}^{N} \sum_{j=1}^{K} y_{ij} \log(\hat{y}_{ij})$$

**Optimizer**
Adam optimization

```
# Create Dense layer with 2 input features and 64 output values
dense1 = Layer_Dense(2, 64)

# Create ReLU activation (to be used with Dense layer):
activation1 = Activation_ReLU()

# Create second Dense layer with 64 input features (as we take output
# of previous layer here) and 3 output values (output values)
dense2 = Layer_Dense(64, 3)

# Create Softmax classifier's combined loss and activation
loss_activation = Activation_Softmax_Loss_CategoricalCrossentropy()

# Create optimizer
optimizer = Optimizer_Adam(learning_rate=0.05, decay=5e-7)
```

**Figure 4: A multimodal presentation supports effective teaching of new material.** A snippet from "Workshop 2: Neural Networks with PyTorch" combines an image of a neural network, its corresponding loss function and mathematical equation, and a Python cell for coding each component of the model—offering multiple representations of the same concept to promote better retention and understanding.

a

b

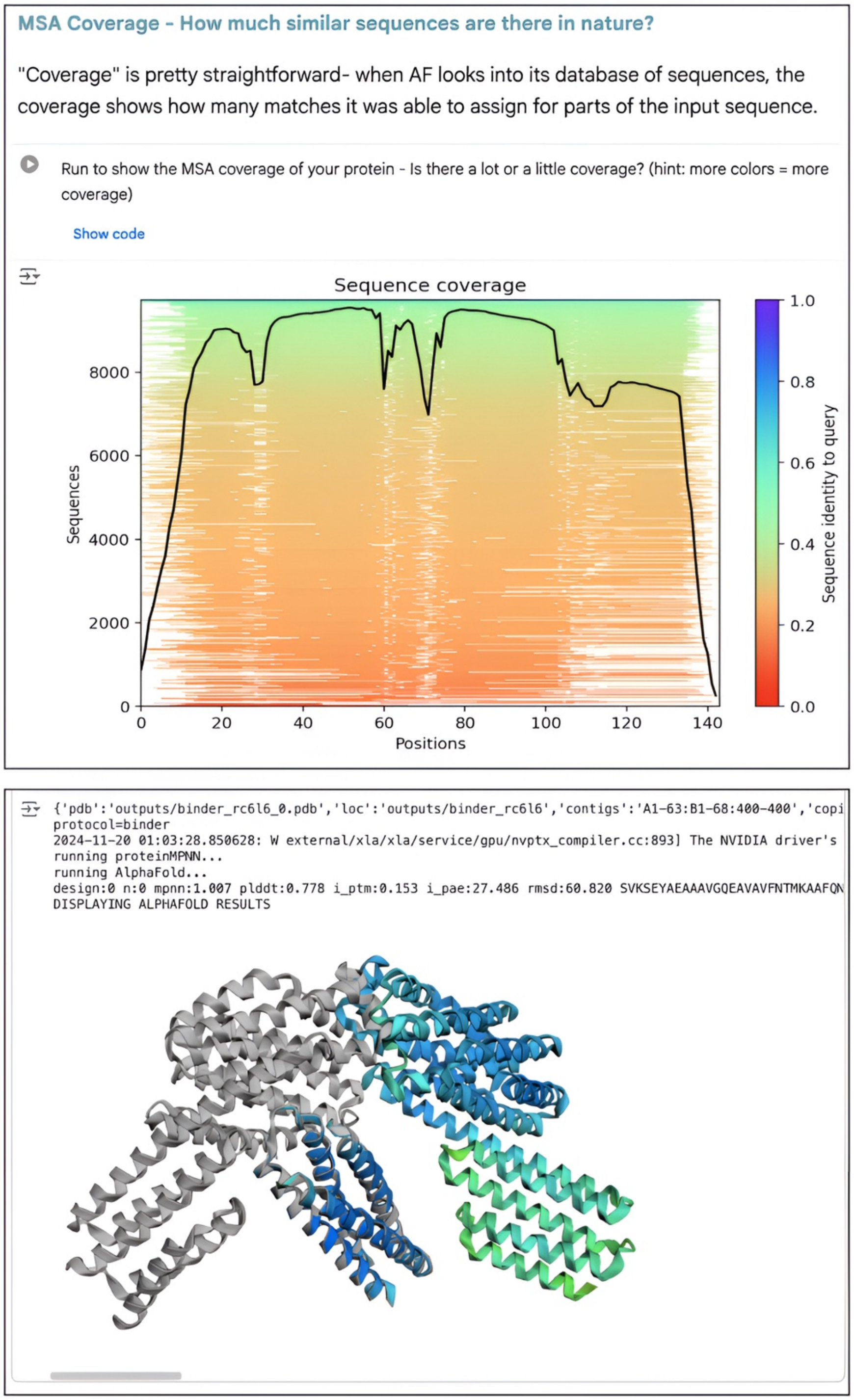

**Figure 5: Interactive visuals provide a stimulating learning experience and help maintain attention.** a) In "Workshop 6: Introduction to AlphaFold", a plot of the MSA coverage is displayed for a protein sequence, and questions follow to guide students on how to interpret this and other metrics provided during the prediction of a protein from AlphaFold2. b) In "Workshop 9: Putting it All Together," a protein is designed by RFDiffusion (gray) and compared to predicted fold from AlphaFold2 (shown in the green-to-blue gradient color scheme). The interactive display allows learners to explore prediction quality with respect to RMSD and to see how different regions vary in confidence based on AlphaFold pLDDT coloring.

## 6 Acknowledgments

We thank Harrison Kinsley for neural network walkthrough videos that inspired the content in WS01; Andrej Karpathy for language model walkthrough videos that inspired the content in WS04; Sergey Ovchinnikov for the RosettaCon 2024 code/tutorial used in WS05, the extended RFDiffusion notebook that inspired the simplified pipeline provided in WS09, and feedback when launching the GitHub repository; the ColabFold[97] developers for the Colab implementation of AlphaFold2 used in WS06; Sergey Lyskov for developing the diffusion model for sinusoidal data used in WS08; the 2024 students of Johns Hopkins 540.614/414 Protein Structure Prediction for testing the first iteration of the notebooks and providing feedback; and Mike Reese and Amy Brusini from the Johns Hopkins Center for Teaching Excellence and Innovation for providing the Instructional Enhancement Grant that supported notebook development.

## 7 Author Contributions

Conception: JJG. Notebook Design – original draft: MFC, GA, PS, CT. Notebook Design – review and editing: MFC, GA, BJC, PS, CT, JJG. Writing – original draft: MFC, GA, BJC. Writing – review and editing: MFC, BJC, JJG, GA, PS. Course instruction: JJG, BJC, MFC.

## 8 Funding

We thank the following funding sources for the support to complete this work; the Johns Hopkins Center for Teaching Excellence and Innovation (CTEI) Instructional Enhancement Grant, NIH R35-GM141881, NSF REU 2244288 and RaMP 2216011, and an NSF Graduate Research Fellow Program to MFC.